\documentclass[prereprint]{revtex4-1}
\pdfoutput=1

\usepackage{graphicx}
\usepackage{bmpsize}
\usepackage{bm}        
\usepackage{hyperref}
\usepackage{amsmath}
\usepackage{amssymb}   
\usepackage{amsopn}
\usepackage{url}
\usepackage{longtable}
\usepackage{color}
\usepackage{listings}
\usepackage[caption=false]{subfig}
\usepackage{filecontents}
\usepackage{lineno}

\IfFileExists{scalerel.sty}
             {
               \usepackage{scalerel}
               \newlength{\heightnu}
               \AtBeginDocument{\settoheight{\heightnu}{$\nu$}}
               
             }
             {
               
             }

\makeatletter
\def\l@paragraph{\@dottedtocline{4}{5.3em}{2.1em}}
\makeatother

\definecolor{bp_red}{RGB}{153, 0, 0}
\definecolor{bp_green}{RGB}{0, 153, 0}
\definecolor{bp_blue}{RGB}{0, 0, 153}

\begin{document}

\title{Physics Potentials of the Hyper-Kamiokande Second Detector in Korea}

\author{Seon-Hee Seo}
\email{sunny.seo@ibs.re.kr}
\affiliation{Center for Underground Physics \\
 Institute for Basic Science \\
 Daejeon, 34126, S. Korea}

\date{\today}

\vglue 1.6cm

\vspace*{2.cm}

\begin{abstract}
Hyper-Kamiokande (Hyper-K) succeeds the very successful Super-K experiment and will consist of a large detector filled with 260~kton purified water
and equipped with 40\% photo-coverage. Physics program of Hyper-K is broad, covering from particle physics to astrophysics and astronomy.
The Hyper-K 1$^{st}$ detector will be built in Japan, and the 2$^{nd}$ detector is considered to be built in Korea
because locating the 2$^{nd}$ detector in Korea improves physics sensitivities in most cases thanks to the longer baseline ($\sim$1,100~km)
and larger overburden ($\sim$1,000~m) for Korean candidate sites.
In this talk, we present overview and physics potentials of the Hyper-K 2$^{nd}$ detector in Korea.
\end{abstract}

\maketitle

\section{Introduction}

\begin{figure}
\begin{center}
\includegraphics[width=0.98\textwidth]{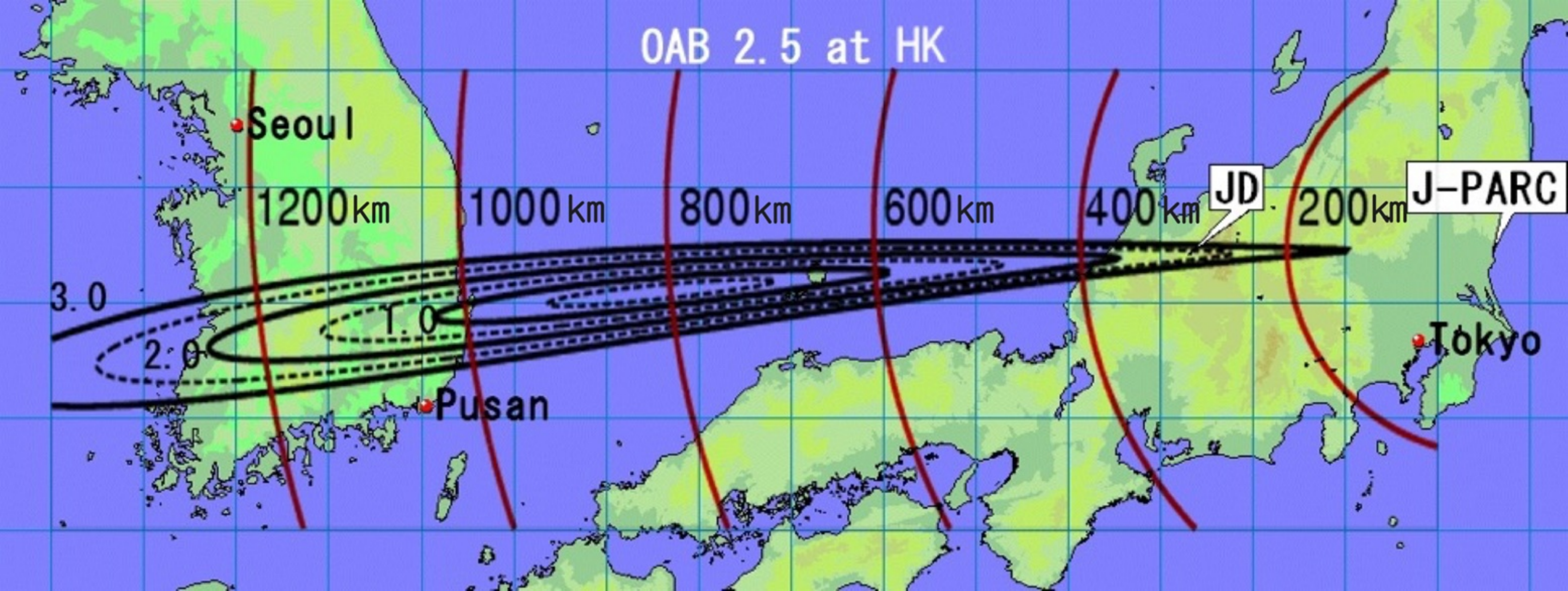}
\end{center}
\caption{Baselines and off-axis angles of the J-PARC neutrino beam in Japan and Korea~\cite{Hagiwara2006,Hagiwara2007}. }
\label{f:OAB}
\end{figure}

Hyper-Kamiokande (Hyper-K) is a next generation water Cherenkov detector consisting of two identical detectors with each 260~kton purified water
and 40\% photo coverage with a combination of 20 inch Hamamatsu PMTs and multi-PMTs (multiple 3 inch PMTs in one enclosure).
Hyper-K will start its construction for the 1$^{st}$ detector~\cite{hk_1st} in April 2020 at Tochibora site where the baseline from J-PARC
is 295\,km with $2.5^\circ$ off-axis angle (OAA).

\begin{figure}[ht]
\centering
\includegraphics[width=0.49\textwidth]{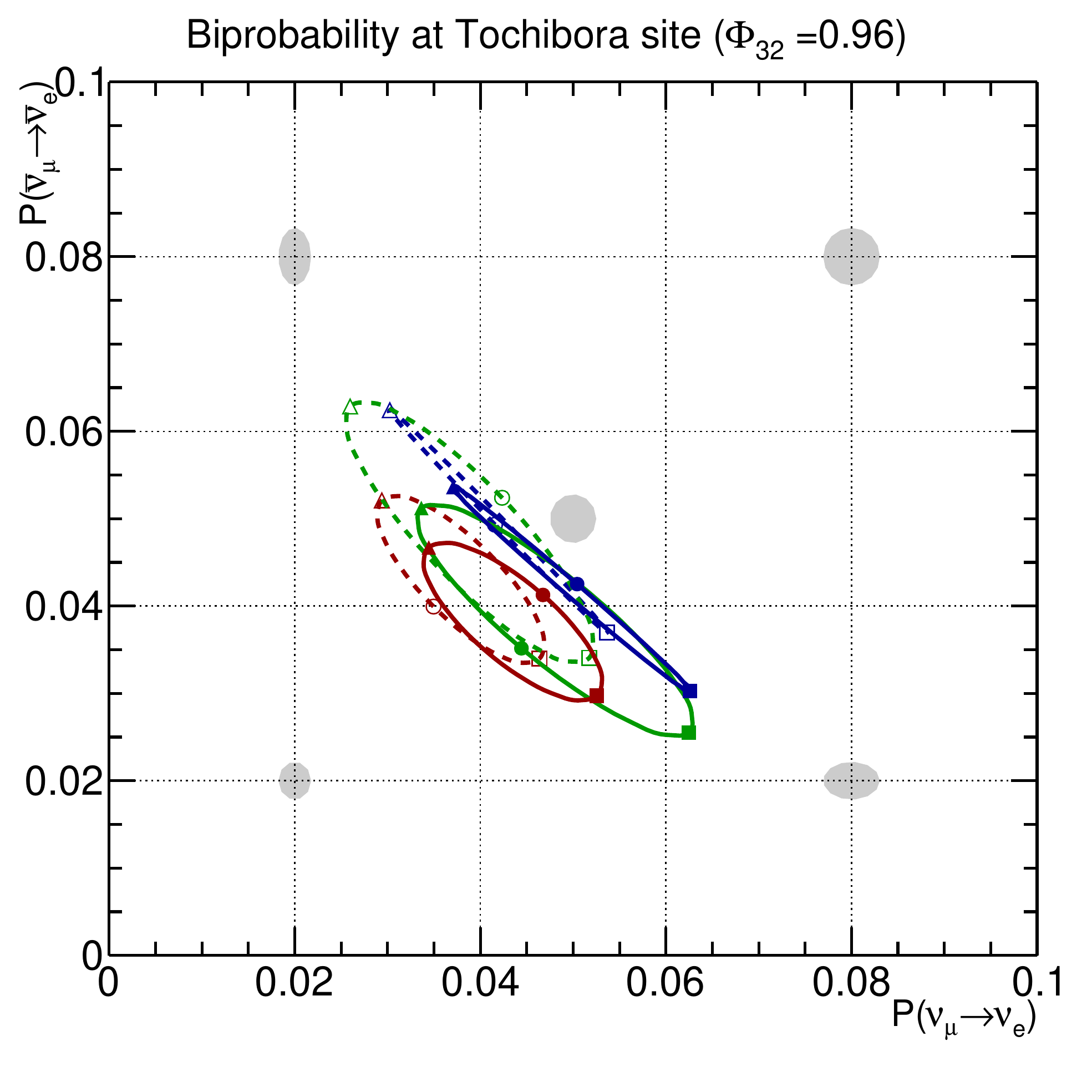}
\includegraphics[width=0.49\textwidth]{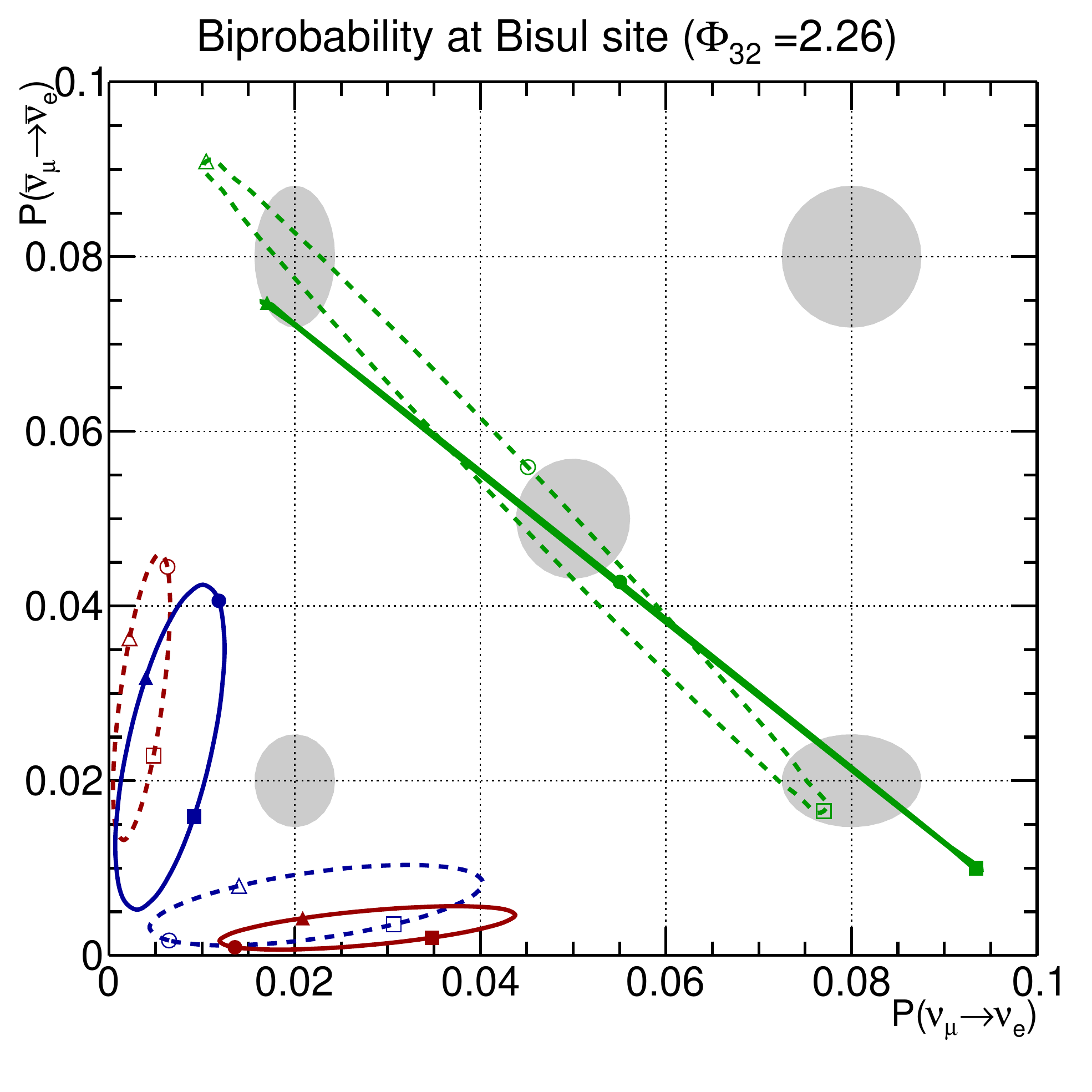}
\begin{minipage}{0.98\textwidth}
\centering
\begin{tabular}{c@{\quad}l@{\qquad}c@{\quad}l@{\qquad}c@{\quad}l@{\qquad}c@{\quad}l}
\textcolor{bp_green}{\rule[.5ex]{3em}{.8pt}} & 520\,MeV
    & \rule[.5ex]{3em}{.8pt} &  $\Delta m^2_{31} > 0$
    &\scalebox{1.0}{$\bullet$}\quad\scalebox{1.05}{$\circ$} & $\delta = 0$
    & \textcolor{bp_green}{\rule[.5ex]{3em}{.8pt}} & 740\,MeV \\

\textcolor{bp_blue}{\rule[.5ex]{3em}{.8pt}} & 620\,MeV
    &\rule[.5ex]{0.6em}{.8pt}\rule[.5ex]{0.6em}{0pt}\rule[.5ex]{0.6em}{.8pt}\rule[.5ex]{0.6em}{0pt}\rule[.5ex]{0.6em}{.8pt} & $\Delta m^2_{31} < 0$
    &\scalebox{1.0}{$\blacktriangle$}\quad\scalebox{1.0}{$\vartriangle$}& $\delta = \pi/2$
    &\textcolor{bp_blue}{\rule[.5ex]{3em}{.8pt}} & 970\,MeV \\

\textcolor{bp_red}{\rule[.5ex]{3em}{.8pt}} & 770\,MeV
    &&
    & $\blacksquare\quad\square$ & $\delta = -\pi/2$
    &\textcolor{bp_red}{\rule[.5ex]{3em}{.8pt}} & 1300\,MeV \\
\end{tabular}
\end{minipage}
\caption{$\nu_{e}$ appearance bi-probabilities at the Hyper-K site in Tochibora, Japan (left) and Mt. Bisul, Korea (right).
Grey ellipses show the sizes of statistical uncertainties for a ten year exposure of one Hyper-K detector,
and $\Phi_{32} \equiv \frac{2}{\pi}\frac{\left|\Delta m^2_{32}\right|L}{4E}$
where $\left|\Delta m^2_{32}\right| = 2.5\times10^{-3}$\,eV$^2$, $E = 620$ ~MeV(left), and 970~MeV (right)~\cite{t2hkk_wp}.
}
\label{f:bp_tochibora}
\end{figure}

The neutrino beam produced at J-PARC currently reaches Korea with $1^{\circ} \sim 3^\circ$ OAA from 1,000\,km to 1,260\,km baselines (see Fig.~\ref{f:OAB}).
This gives an opportunity to locate the 2$^{nd}$ detector in Korea since the longer baseline would benefit various physics potentials
such as precise measurement of CP violation phase and neutrino mass ordering determination.
Bi-probability plots in Fig.~\ref{f:bp_tochibora} clearly shows that
it is easier to determine neutrino mass ordering and CP violation phase in a Korean Mt. Bisul site.
This is due to the longer baseline and higher energy reach from the 1$^{st}$ and 2$^{nd}$ oscillation maxima.

Table~\ref{t:six_sites} shows six possible candidate sites in Korea.
Thanks to larger overburden of $\sim$1000~m in Korean candidate sites than Japan Tochibora site (650~m),
low energy physics sensitivities such as solar neutrinos, supernova relic neutrinos are expected to be improved.

\begin{table}[ht]
\small
\caption{Six candidate sites of the Hyper-K 2$^{nd}$ detector in Korea with off-axis angles between 1$^{\circ}$ and 2.5$^{\circ}$.
The baseline is the distance from the production point of the J-PARC neutrino beam to the candidate sites~\cite{t2hkk_wp}.}
\centering
\begin{tabular*}{0.9\textwidth}{@{\extracolsep{\fill}} l c c c l}

  \hline \hline
Site           & Off-axis & Baseline & Height
               &    Composition of rock \\[-1.ex]
               & angle    & (km)     & (m) & \\
\hline
 Mt. Bisul     & 1.3$^{\circ}$   &  1,088    & 1,084
               &    Granite porphyry,\\[-1.ex] 
               &&&& andesitic breccia \\[0.4ex]
 Mt. Hwangmae  & 1.9$^{\circ}$   &  1,141   & 1,113
        &    Flake granite, \\[-1.ex] 
               &&&& porphyritic gneiss  \\[0.4ex]
 Mt. Sambong   & 2.1$^{\circ}$   &  1,169   & 1,186
               &    Porphyritic granite, \\[-1.ex] 
               &&&& biotite gneiss \\[0.4ex]
Mt. Unjang    & 2.2$^{\circ}$   &  1,190   & 1,125
               &    Rhyolite, granite porphyry, \\[-1.2ex] 
               &&&& quartz porphyry  \\ [0.4ex]
 Mt. Bohyun    & 2.3$^{\circ}$  &  1,043   & 1,124
        &    Granite, volcanic rocks, \\[-1.ex] 
               &&&& volcanic breccia \\[0.4ex]
 Mt. Minjuji   & 2.4$^{\circ}$   &  1,145  & 1,242
               &    Granite, biotite gneiss \\
  \hline \hline
\end{tabular*}
\label{t:six_sites}
\end{table}

Due to a very limited space here we present only some physics sensitivities, but more details are found in ~\cite{t2hkk_wp}.

\section{Physics potentials}

Sensitivity studies are performed using NEUT~\cite{Hayato:2009zz} 5.3.2 for neutrino interaction generator and full simulation of Super-K scaled to Hyper-K
for the expected event rates,
Prob3++~\cite{prob3++} for the oscillation probabilities, and a constant matter density of 3.0~g/cm$^2$ for a 1100~km baseline
with 1.5$^{\circ}$, 2.0$^{\circ}$, or 2.5$^{\circ}$ OAA for hypothetical Korean sites.

In our sensitivity studies, unless otherwise specified, 10 years of operation with 1.3~MW beam power is assumed with 1:3 ratio of neutrino to antineutrino modes,
and this corresponds to $2.7\times10^{22}$ proton on target (POT).
A reasonable systematic uncertainty model is also applied ~\cite{t2hkk_wp}.

According to our sensitivity studies, when the CP is maximally violated with known mass ordering, there is almost no difference in the CP violation sensitivity
between Japan Detector (JD) plus Korean Detector (KD) configuration and two Japan Detectors (JD$\times$2).
However, when the CP is a little non-maximally violated JD$+$KD sensitivity is better.

\begin {figure}[htbp]
\begin{center}
\includegraphics[width=0.49\textwidth]{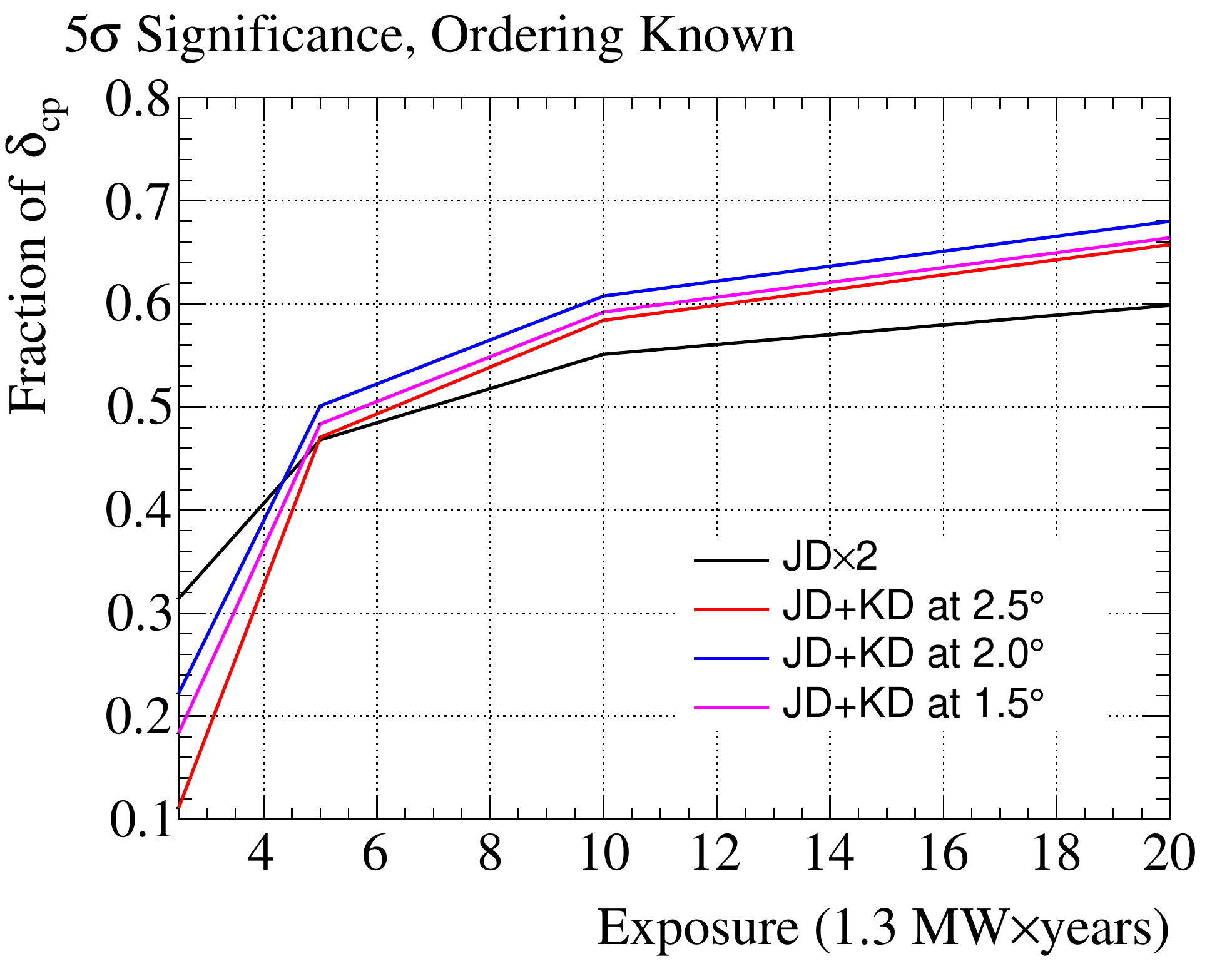}
\includegraphics[width=0.49\textwidth]{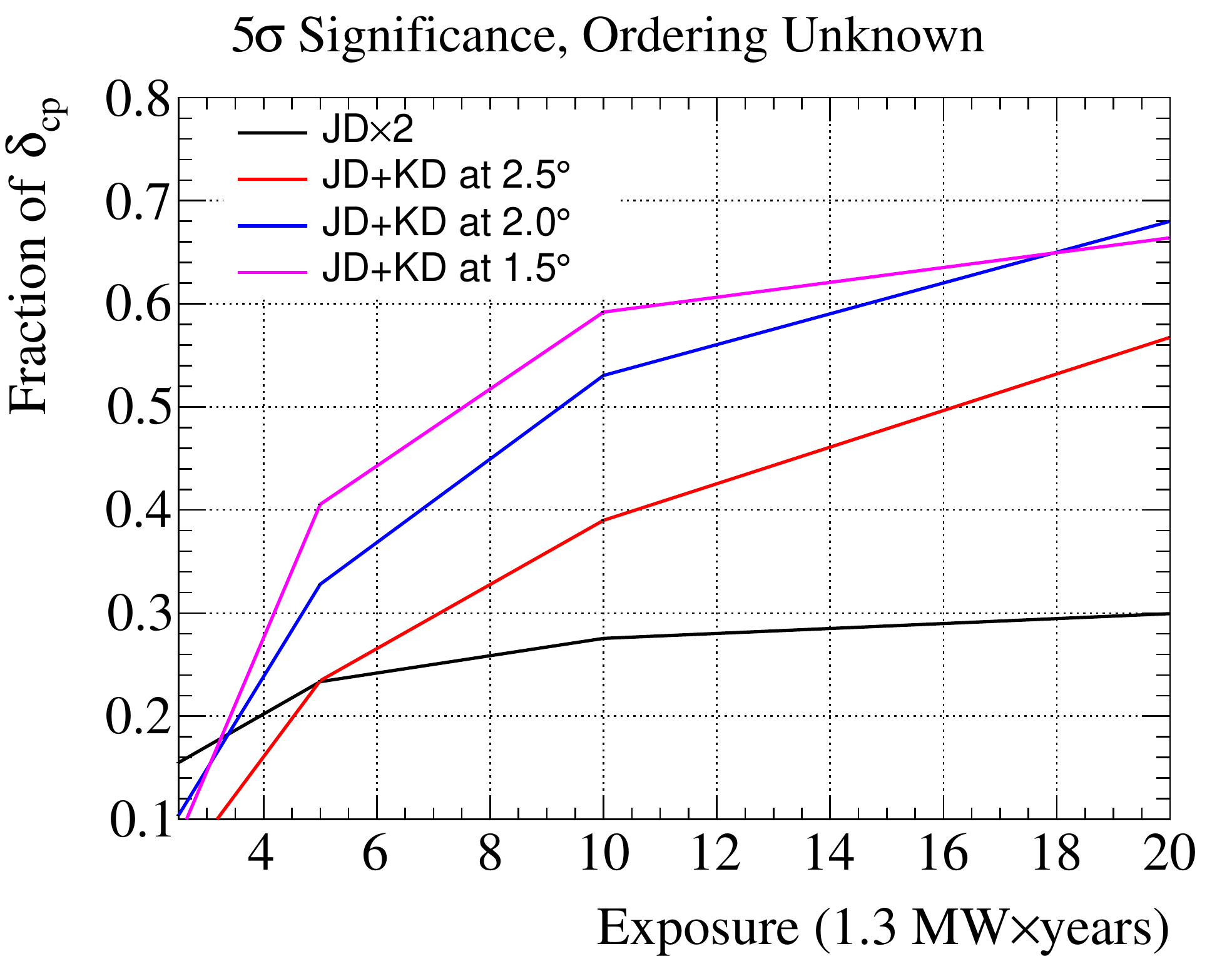}
\caption{The fraction of $\delta_{cp}$ values (averaging over the true mass ordering)
to reject the CP conserving values of $\delta_{cp}$ with at least a 5$\sigma$ significance for known (left) and unknown (right) mass ordering
~\cite{t2hkk_wp}. }
\label{fig:cpv_exp}
\end{center}
\end {figure}

Figure~\ref{fig:cpv_exp} shows the progress of $\delta_{cp}$ violation discovery potential (minimum 5$\sigma$ significance) with exposure
for known (left) and unknown (right) mass ordering.
For the known mass ordering case, due to the limited statistics, the sensitivity of the first 4.5 years is worse for any JD$+$KD configuration than JD$\times$2.
After then any JD$+$KD configuration sensitivity gets better and reaches close to $\sim$60\% fraction of $\delta_{cp}$,
while $\sim$55\% for JD$\times$2 limited by systematic uncertainties.

\begin {figure}[htbp]
\begin{center}
\includegraphics[width=0.49\textwidth]{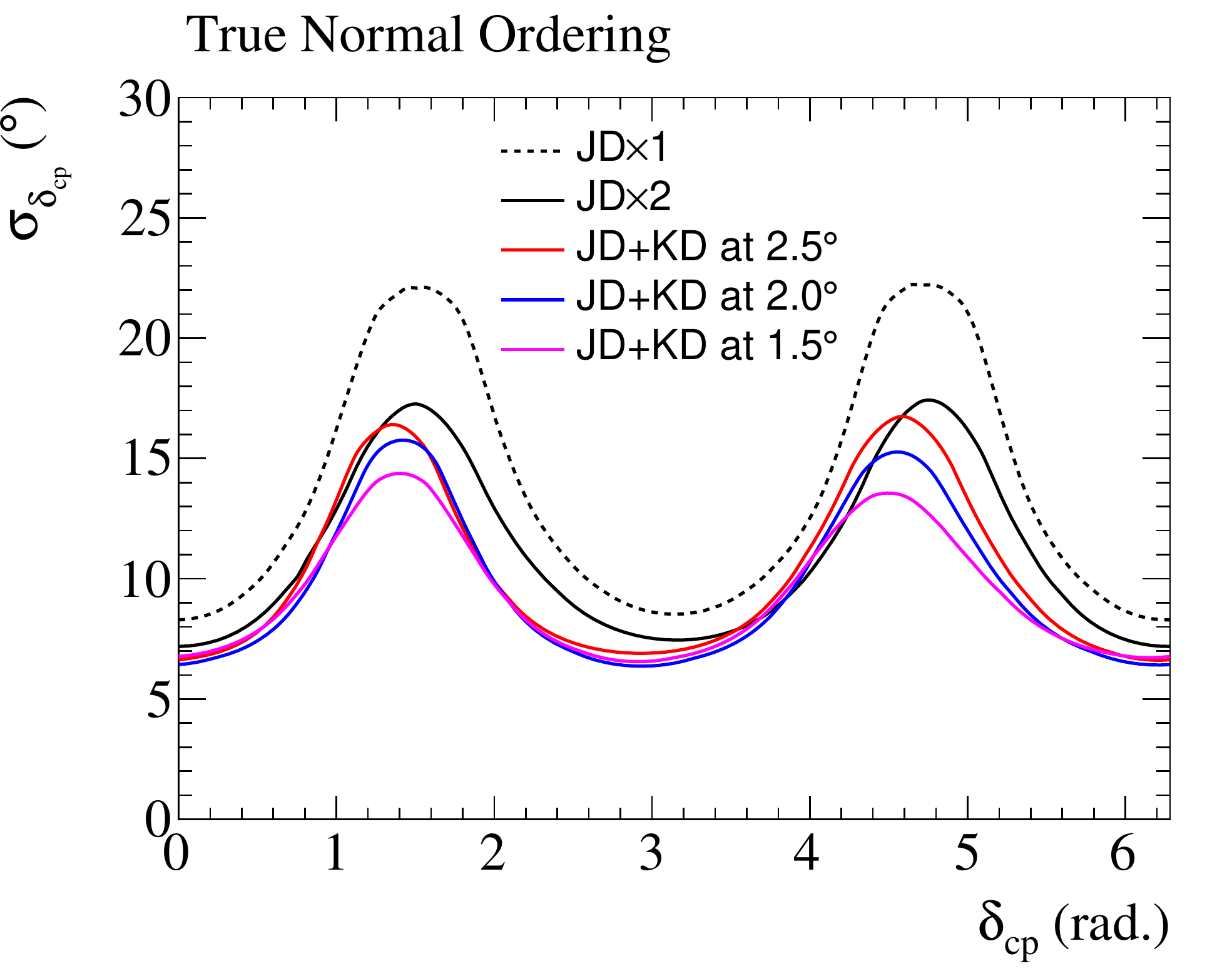}
\includegraphics[width=0.49\textwidth]{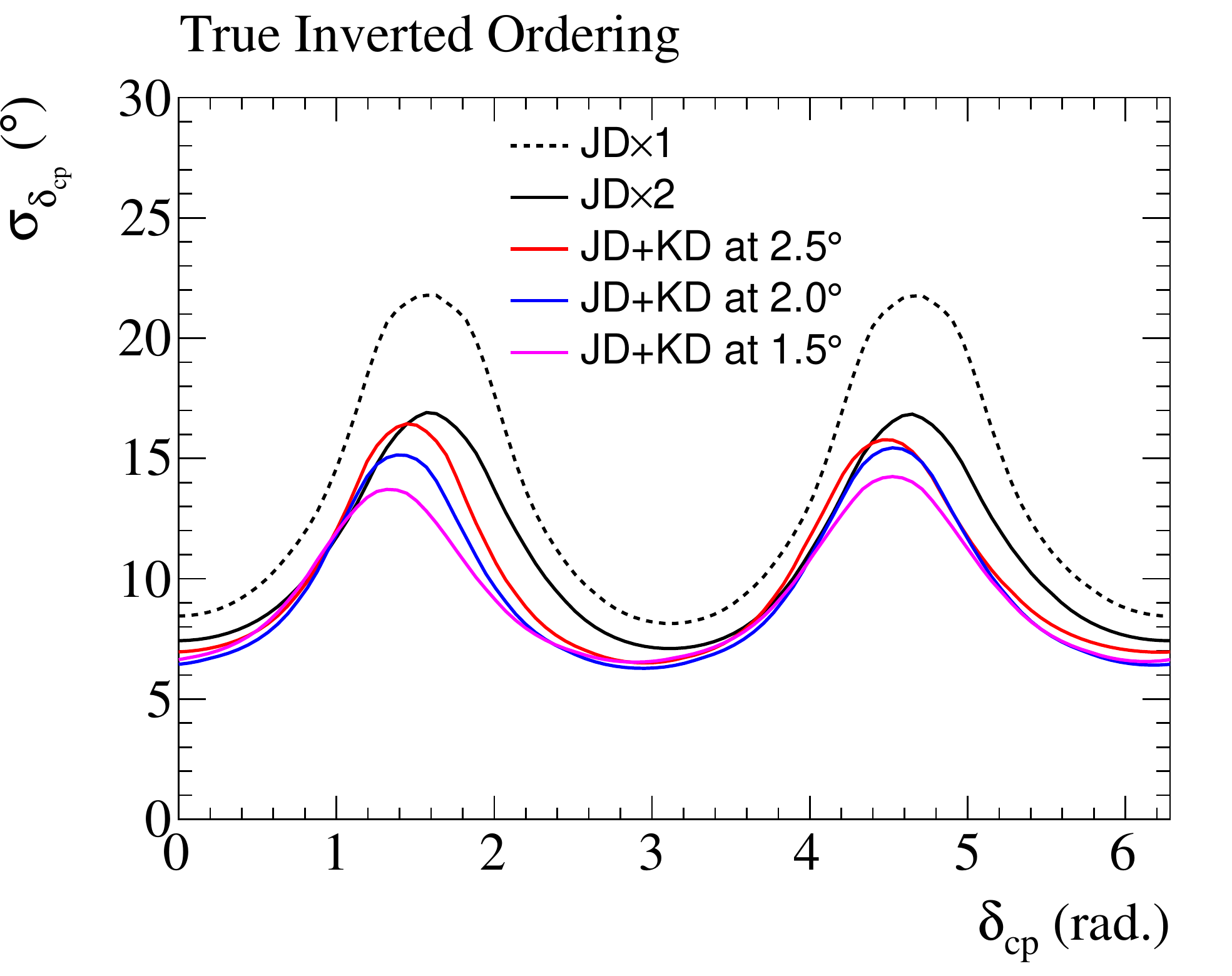}
\caption{The 1$\sigma$ precision of the $\delta_{cp}$ measurement as a function of the true $\delta_{cp}$ value with unknown mass ordering~\cite{t2hkk_wp}.}.
\label{fig:cp_precision}
\end{center}
\end {figure}

The aim of Hyper-K is not only to determine CP violation or not but also precise measurement of $\delta_{cp}$ value.
By locating the 2$^{nd}$ detector in Korea we can improve this sensitivity and this is an important input for flavor symmetry models.
Figure~\ref{fig:cp_precision} shows the 1$\sigma$ precision of the $\delta_{cp}$ measurement as a function of the true $\delta_{cp}$ value
with the mass ordering unknown.
When the CP is maximally violated cases, the precision changes from $\sim 22^{\circ}$ (JD$\times$1) and $\sim 17^{\circ}$ (JD$\times$2)
to $13\sim14^{\circ}$ (JD$+$KD at 1.5$^\circ$ OAA).

\begin {figure}[ht]
\begin{center}
\includegraphics[width=0.75\textwidth]{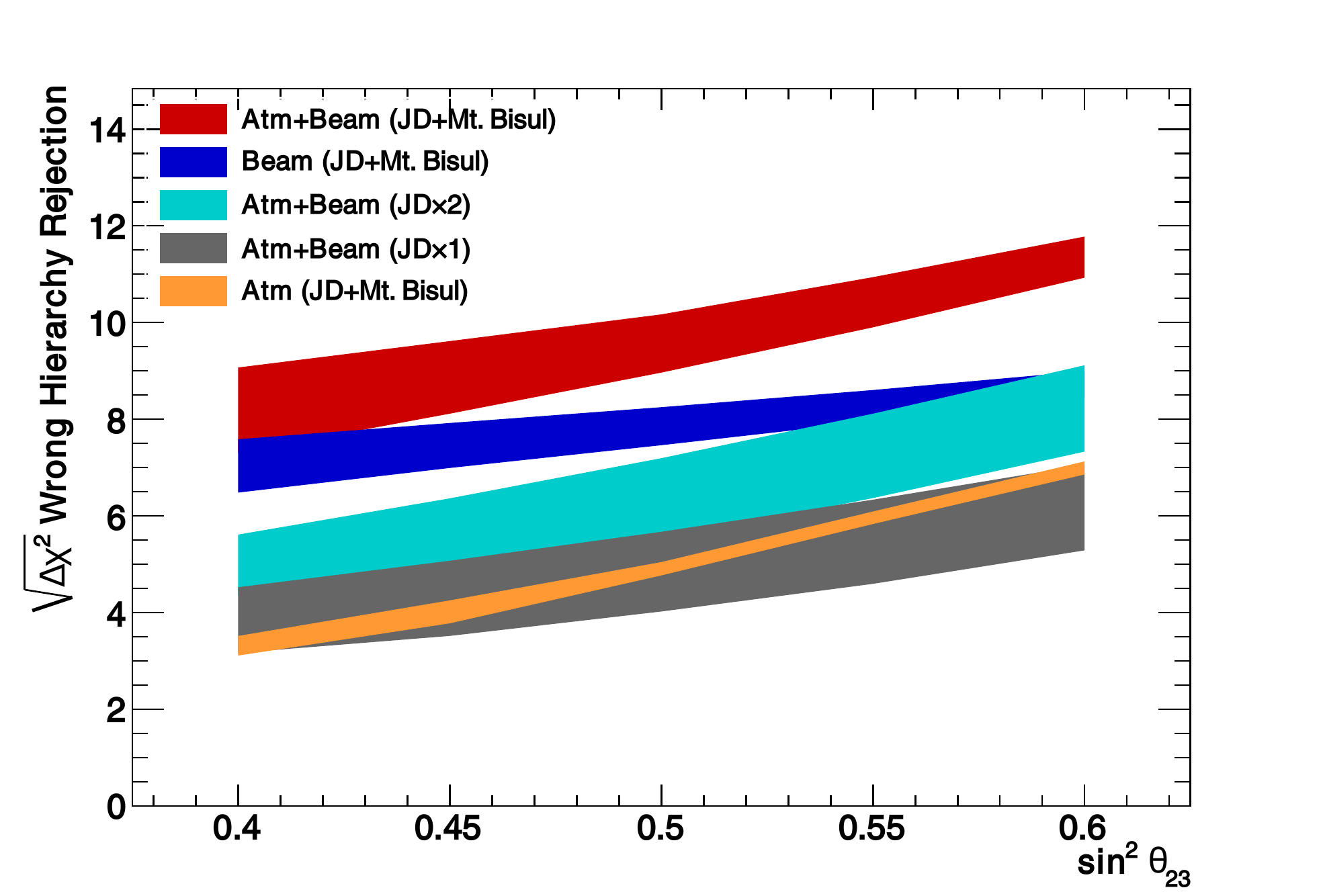}
\caption{Sensitivity to reject wrong mass ordering assuming true normal ordering
             with a combined measurement of beam and atmospheric neutrinos for a 10 year exposure.
             The x-axis shows the possible ranges of $\sin^{2} \theta_{23}$ and the
             width of the bands comes from the variation in sensitivity with $\delta_{cp}$~\cite{t2hkk_wp}. }
\label{fig:jdkd_hier_10}
\end{center}
\end{figure}

Sensitivity to determine mass ordering is improved with JD$+$KD configuration.
Atmospheric neutrinos add additional increase of the sensitivity.
Figure~\ref{fig:jdkd_hier_10} shows wrong mass ordering rejection significance as a function of different octant values
for true normal ordering case.
For any $\sin^{2} \theta_{23}$ value between 0.4 and 0.6, more than 8$\sigma$ sensitivity is expected
for JD$+$KD (Mt. Bisul) configuration when the atmospheric neutrinos are combined with beam neutrinos.

\section{Conclusion and Prospects}
We have performed sensitivity studies for various cases of locating the 2$^{nd}$ detector in Korea (1.5$^{\circ}$, 2.0$^{\circ}$, or 2.5$^{\circ}$ OAA at 1,100~km)
versus two detectors in Japan (2.5$^{\circ}$ OAA at 295~km).
In most cases the sensitivities are improved and atmospheric neutrinos add additional sensitivities in all cases.
We find that a smaller OAA site in Korea gives better sensitivities in most cases.
Therefore the Mt. Bisul site (1.3$^{\circ}$ OAA, 1,088~km baseline) is the strongest candidate for the 2$^{nd}$ detector in Korea.
Further design studies are needed for the 2$^{nd}$ detector to maximize physics potentials, for example, by enlarging the detector size
and/or adding Gadolinium to tag neutrons.

\end{document}